\documentclass[twocolumn,floatfix,prl]{revtex4-2}%

\usepackage[dvipdfmx]{graphicx}%
\usepackage{amsmath}%
\usepackage{amsfonts}%
\usepackage{amssymb}
\usepackage{bm}
\usepackage{color}
\usepackage{ulem}
\usepackage{comment}
\usepackage{braket}
\usepackage{color}

\def\q{{ {\bm q} }}

\def\0{{ {\bm 0} }}

\allowdisplaybreaks[4]

\begin{document}
\title{
Development of spin fluctuations 
under the presence of $d$-wave bond order in 
cuprate superconductors
}
\author{Satoshi Ando, Youichi Yamakawa, Seiichiro Onari, Hiroshi Kontani}
\date{\today }
\begin{abstract}

In cuprate superconductors, 
superconductivity appears below  the CDW transition temperature $T_{\rm CDW}$.
However, many-body electronic states under the CDW order
are still far from understood.
Here, we study the development of the spin fluctuations
under the presence of $d$-wave bond order (BO) with 
wavevector $\q=(\pi/2,0),(0,\pi/2)$,
which is derived from the paramagnon interference mechanism
in recent theoretical studies.
Based on the $4\times1$ and $4\times4$ cluster Hubbard models,
 the feedback effects between spin susceptibility and 
self-energy are calculated self-consistently
by using the fluctuation-exchange (FLEX) approximation.
 It is found that the $d$-wave BO
leads to a sizable suppression of the nuclear magnetic relaxation rate $1/T_1$.
In contrast, the reduction in $T_{\rm c}$
 is small,
since the static susceptibility $\chi^s({\bm Q}_s)$ is affected by the BO just slightly.
It is verified that the $d$-wave BO scenario is consistent with 
the experimental electronic properties below $T_{\rm CDW}$.
	
\end{abstract}
\address{
Department of Physics,
Nagoya University,
Nagoya 464-8602,
Japan
}
\sloppy
\maketitle



\begin{figure}[!htb]
\includegraphics[width=.7\linewidth]{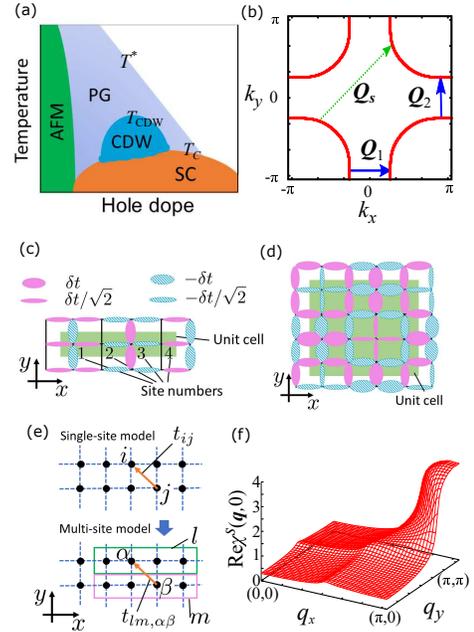}
\caption{
(a) Schematic phase diagram of YBCO.
(b) Original FS. $\bm{Q}_1$ and $\bm{Q}_2$ are the axial nesting vectors related to the period of the BO. $\bm{Q}_s=(\pi,\pi)$ is a nesting vector.
(c) Schematic single-$\bm{q}$ four-period $d$-wave BO. $\delta t$ is the modulation of the hopping.  
(d) Schematic double-$\bm{q}$ $4\times4$ period $d$-wave BO.
(e) Relation between indices $i,j$ and $\alpha,\beta,l,m$. 
$i(j)$ is the site index over all the space.
$\alpha(\beta)$ is the site index in a unit cell $l(m)$.
(f) Spin susceptibility without the BO at $T=0.01$. It has a peak at $\bm{Q}_s$. 
\label{fig:intro}
}
\end{figure}

The electronic states in cuprate superconductors
have been studied as central issues in condensed matter physics.
Especially, the mechanisms and the true order parameters of 
the charge density wave (CDW) state and the 
pseudogap state have intensively studied  in the last decade.
Figure \ref{fig:intro}(a) shows a schematic phase diagram of 
Y-based compound (YBCO).
The antiferromagnetic (AFM) order at filling $n=1$ 
rapidly disappears with hole doping.
For wide doping range, the pseudogap phase appears below $T^*$.
As discussed in Refs. \cite{D.Senechal,B. Kyung,T.A.Maier},
large quasiparticle damping could induce sizable reduction in the density of states (DOS).
On the other hand,
the rotational symmetry breaking \cite{A.Shekhter,R.-H.He,Y.Sato}
and the time reversal symmetry breaking \cite{loopcurrent} have been observed recently.
To explain them,  nematic transitions such as $d$-wave bond order (BO) \cite{S-Kawaguchi-Cu,S-Tsuchiizu-Cu,Y.Yamakawa-Cu} and charge or spin-loop-current \cite{spin-loop-Kontani,C.M.Varma,I.Affleck,F.C.Zhang,R.Tazai-chargeloop} are proposed. 

Inside the pseudogap region, the CDW-type order emerges below $T_{\rm CDW}$,
higher than the 
superconducting transition temperature $T_{\rm c}$ \cite{G.Ghiringhelli, J.Chang, E.Blackburn,M.H2011,dynamical,E.H.Shilva, W.Tabis, R.Comin,M.H,S.Blanco,suppressTc, R.Comin2015, R.Comin2014,W.Tabis2017,K.Fujita-Cu-STM,T.Hanagri,Y.Kohsaka,M.J.Lawler,3DCDWsuppress,fluctuation-Cu}.  
     Spin-fluctuation scenarios  \cite{J.C.Davis,M.A.Metlitski,C.Husemann,K.B.Efetov,S.Sachdev,V.Mishra} and superconducting-fluctuation scenarios \cite{E.Berg,P.A.Lee,E.Fradkin,Y.Wang} have been proposed as its driving mechanism. 
Its wavevector is ${\bm q}=(\delta,0),\ (0,\delta)$
with $\delta\approx \pi/2$
according to the X-ray scattering experiments 
\cite{G.Ghiringhelli, J.Chang, E.Blackburn,M.H2011,dynamical,E.H.Shilva, W.Tabis, R.Comin,M.H,S.Blanco,suppressTc, R.Comin2015, R.Comin2014,W.Tabis2017,fluctuation-Cu} 
and the STM measurements 
\cite{K.Fujita-Cu-STM,T.Hanagri,Y.Kohsaka,M.J.Lawler}. 
Conventional CDW with local charge density modulation
$\Delta n_i$ ($\equiv n_i-n_i^0$) is suppressed by 
the on-site Coulomb interaction $U$. In contrast, off-site order parameters, 
such as the BO parameter $\delta t_{ij}$ 
($\propto \braket{ c_{i\sigma}^\dag c_{j\sigma} } 
- \braket{ c_{i\sigma}^\dag c_{j\sigma} }_0$) 
are not suppressed by $U$.
In fact, various multipole orders are induced by
the interference mechanism between paramagnons, 
as revealed by the density wave (DW) equation 
\cite{S-Kawaguchi-Cu,Y.Yamakawa-Cu,S-Onari-form,S.onari2012,Y.Yamakawa2016,R.Tazai2020,S.Onari2019,S.Onari2020},  
the functional renormalization group (fRG) studies  
\cite{S-Tsuchiizu-Cu,W.Metzner,C.Bourbonnais,M.Tsuchiizu2013,R.Tazai2016,C.Honerkamp}, and other theories \cite{Y.Wang2014,tazai2019}.
In Refs. \cite{S-Kawaguchi-Cu,S-Tsuchiizu-Cu}, it was predicted that the nematicity in the pseudogap-phase below $T^*$ 
 is the ferro ($\bm{q}=\bm{0}$) $d$-wave BO, and the experimental CDW phase is the antiferro ($\bm{q}=(\delta,0)$) $d$-wave BO. 

In many cuprate superconductors, 
superconductivity appears under the presence of the CDW-type order.
However, many-body electronic states under the CDW-type order
are still far from understood.
For example, the Hall and Seebeck coefficients show large positive values for $T\gg T_\mathrm{CDW}$ \cite{H.Kontani,H.Kontani-Fermi-liquid}, while they suddenly decrease and become negative below $T_\mathrm{CDW}$ 
\cite{hall-seebeck,hall,seebeck}.
These observations indicate the Fermi pocket formation 
due to the band folding by the antiferro BO.
Also, the nuclear magnetic relaxation rate $1/T_1$,
which represents the low-energy spin-fluctuation strength,
is strongly suppressed below $T_\mathrm{CDW}$ \cite{M.Takigawa-Cu-T1T}.   
In contrast, 
$T_c$ is suppressed by the CDW-type order only slightly in YBCO
\cite{W.Tabis,M.H,S.Blanco},
and the finite-energy spin-fluctuation strength (at $E\sim30$ meV)
grows monotonically even below $T_\mathrm{CDW}$ 
according to the neutron inelastic scattering experiments
\cite{neutron}.

In this paper,
we study the development of the spin fluctuations
under the presence of antiferro $d$-wave BO.
Based on the $4\times1$ ($4\times4$) cluster Hubbard model
with single-$\bm q$ (double-${\bm q}$) BO parameters,
we study the feedback effects between spin-fluctuations and 
self-energy  self-consistently
 by using the fluctuation-exchange (FLEX) approximation \cite{FLEX1}.
The pseudogap formation due to the $d$-wave BO
leads to the sizable suppression of $1/T_1T$.
In contrast, the reduction in $T_{\rm c}$
due to the BO remain small,
since $\chi^s({\bm Q}_s)$ is affected by the BO just slightly.
It is found that the $d$-wave BO scenario naturally explains essential electronic properties below $T_{\rm CDW}$ in Fig \ref{fig:intro}(a).

The present study reveals the importance of the site dependent 
electronic states under the BO state, because 
the site averaging approximation (SAA) leads to inappropriate results as we will show later.
Therefore, analysis of the cluster Hubbard model
is necessary for understanding the electronic states in the BO phase.



The Hubbard Hamiltonian is represented by $H=H_0+H_{U}$, 
where the hopping term is 
$H_0=\sum_\sigma \sum_{ij}t_{ij}a^\dag_{i\sigma}a_{j\sigma}$, and the interaction term is $H_{U}=\sum_{i} U a^\dag_{i\downarrow}a^\dag_{i\uparrow}a_{i\uparrow}a_{i\downarrow}$.
 Here, $a^\dag_{i\sigma}$ and  $a_{i\sigma}$ are the creation and annihilation operators of the electron on site $i$ with spin $\sigma$, respectively. $t_{ij}$ is the transfer integral from site $j$ to $i$.
 We define $t_1$, $t_2$, and $t_3$ as the first, the second, and the third nearest neighbor hopping without any orders, respectively.
Here, we set  $t_1=-1$, $t_2=1/6$, and $t_3=-1/5$.

 Figure \ref{fig:intro}(b) shows the Fermi surface (FS) without the $d$-wave BO. The nesting vector $\bm{Q}_s=(\pi,\pi)$ gives the peak of the spin fluctuation.  $\bm{Q}_1=(\delta,0)$ and $\bm{Q}_2=(0,\delta)$ are the axial nesting vectors related to the period of the BO.
 We set $\delta=\pi/2$ and $n=1.015$ to discuss the four-period order. 
 (We also investigated the five-period ordered case at $n=0.888$, and confirmed that the results are qualitatively similar to those in the four-period ordered case.) 
 
     The antiferro $d$-wave BO is represented by the modulation of the hopping as
\begin{align}
\delta t^{\bm{q}}_{ij} = &\delta t\left[(\delta_{\bm{r}_i-\bm{r}_j,\hat{\bm{x}}}-\delta_{\bm{r}_i-\bm{r}_j,\hat{\bm{y}}})+(i\leftrightarrow j)\right] \nonumber\\ 
&\times\mathrm{Re}\left[\mathrm{e}^{-\mathrm{i}\left(\bm{q}\cdot\frac{\bm{r}_i+\bm{r}_j}{2}-\gamma \right)}\right],
\label{equ:BO}
\end{align} 
where $\delta t^{\bm{q}}_{ij}$ has an opposite sign for $x$-direction and $y$-direction.  The period and direction of the modulation are decided by wave number $\bm{q}$.
$\delta t^{\bm{Q}_1}_{ij}$ is shown in Fig. \ref{fig:intro}(c).
$\bm{r}_i$ is the position of site $i$, and $\hat{\bm{x}}$ ($\hat{\bm{y}}$) is the $x$ ($y$)-direction unit vector. 
$\delta t$ is the BO strength, and we assume that $\delta t$ has  the BCS-like $T$ dependence, $\delta t(T)=\delta t_0\mathrm{tanh}(1.74\sqrt{T_{\mathrm{BO}}/T-1})$.   $T_{\mathrm{BO}}$ is the BO transition temperature, which corresponds to $T_{\mathrm{CDW}}$.
$\gamma$ is the phase parameter of the BO.
We have confirmed that  both  $1/T_1T$ and $T_c$ are independent of $\gamma$ as shown in Figs. \ref{fig:sp1}(e) and (f) in the supplemental material (SM), so we set $\gamma=0$ in this study.      
 We call $\delta t^{\bm{Q}_1}_{ij}(T)$ in Eq. (\ref{equ:BO})  the ``single-$\bm{q}$'' BO.

  We also discuss the ``double-$\bm{q}$'' BO represented by $\delta t^{{\bm{Q}_1}}_{ij}(T)+\delta t^{{\bm{Q}_2}}_{ij}(T)$. 
  Figure \ref{fig:intro}(d) shows the schematic picture of $4\times4$ period double-$\bm{q}$ BO with ${\bm{Q}_1}=(\pi/2,0)$ and ${\bm{Q}_2}=(0,\pi/2)$.  Hereafter,  we set $\delta t_0=0.075$ and $T_{\mathrm{BO}}=0.02$ to   
  follow the BCS universal ratio between the gap size and the transition temperature as we will explain later. 
 From $|t_1|\simeq0.5\mathrm{eV}$ given by the first principle calculation \cite{P.Hansmann}, $T_{\mathrm{BO}}=0.02\simeq120\mathrm{K}$ is estimated.   

      We introduce the cluster-Hubbard models to consider the electronic states under the antiferro BO. The hopping  Hamiltonian $H_0$ can be rewritten by     
\begin{align}
H_0=\sum_\sigma \sum_{lm\alpha \beta}t_{lm,\alpha \beta}a^\dag_{l\alpha \sigma}a_{m\beta \sigma},
\label{equ:hamiltonian_orderd}    
\end{align}
where $\alpha(\beta)$ is the site index in a unit cell $l(m)$. As shown in Fig. \ref{fig:intro}(e), a set of $l$ and $\alpha$ ($m$ and $\beta$) corresponds to a site $i(j)$.

We use the site-dependent FLEX approximation to calculate the spin susceptibility 
$\hat{\chi}^s(q)=\hat{\chi}^0(q)\left[1-U\hat{\chi}^0(q)\right]^{-1}$ in the cluster models, where $\hat{\chi}^0(q)$ is the irreducible susceptibility.
$\hat{\chi}^s(q)$ and $\hat{\chi}^0(q)$ are $4\times4$ matrices for the single-$\bm{q}$ ($16\times16$ matrices for the double-$\bm{q}$) case.
We use the short notations $k=(\bm{k},\mathrm{i}\varepsilon_n)$ and $q=(\bm{q},\mathrm{i}\omega_l)$, where $\varepsilon_n$ and $\omega_l$ are
fermion and boson Matsubara frequencies, respectively. 
 We ignore the modulation in the FS shape by the FLEX self-energy, approximately, as we explain in SM. A. 
 
Here, we explain the ``unfolding'' procedure \cite{unfold}.
 A physical quantity under the antiferro order in Fig. 1 (e) is
expressed as $A(\bm{r}_i-\bm{r}_j) = A_{\alpha\beta}(\bm{R}_l-\bm{R}_m)$, and its Fourier transformation is $A_{\alpha\beta}(\bm{k})=
\sum_l A_{\alpha \beta}(\bm{R}_l)\exp{\{-\mathrm{i}(\bm{k}\cdot\bm{R}_l)\}}$.
Here, $\bm{R}_l$ is the position of the unit cell $l$, and we put 
$-\pi<k_{x,y}\leq\pi$. Then, its average is given as
\begin{align}
A(\bm{k})=
\frac{1}{N}_c\sum_{\alpha\beta} A_{\alpha \beta}(\bm{k})
 \mathrm{e}^{-\mathrm{i}\bm{k} \cdot (\bm{\tau}_\alpha-\bm{\tau}_\beta)},
\label{equ:unfold}
\end{align}
where $N_{c}$ is the number of the sites in a unit cell, and  $\bm{\tau}_{\alpha}$ 
is the position of the site $\alpha$.
The obtained $A(\bm{k})$ is unfolded to the single-site Brillouin zone because the translational symmetry is recovered approximately.

 Figure \ref{fig:intro}(f) shows the real part of the  spin susceptibility Re$\chi^s(\bm{q},0)$ without the BO at $T=0.01$ for $U=4.75$.   
Re$\chi^s(\bm{q},0)$ has the peak at $\bm{Q}_s=(\pi,\pi)$ due to the nesting shown in Fig. \ref{fig:intro}(b).  

     Figure \ref{fig:DOS,Fermi}(a) shows the $T$ dependence of the DOS under the single-$\bm{q}$ BO without the FLEX self-energy. We estimate the bare (: without the FLEX self-energy) gap size $2\Delta_\mathrm{BO}$ from the width between the peaks of the DOS.
      Note that the relation between $\Delta_\mathrm{BO}(T)$ and $\delta t (T)$ is $2\Delta_\mathrm{BO}(T)\simeq4\delta t (T)$ under the single-$\bm{q}$ BO.
       {(The DOS under $s$-wave and $s'$-wave orders is discussed in SM. D.)}
     This relation can be understood from the inter-band hybridization due to the band holding \cite{R.Tazai2020,reconst-FS} (see SM. B.).      
In Fig. \ref{fig:DOS,Fermi}(b), we compare the DOS under the single-$\bm{q}$ BO with the DOS under the double-$\bm{q}$ one. The decrease in the DOS
is larger in the double-$\bm{q}$ case.
  To compare with experiments, we calculate the renormalized gap as $2\Delta^*_{\mathrm{BO}}(\bm{k},T)=2\Delta_{\mathrm{BO}}(T)/Z(\bm{k},T)$, where
$Z(\bm{k},T)=\left[1-\frac{\Sigma(\bm{k},\mathrm{i}\pi T)-\Sigma(\bm{k},-\mathrm{i}\pi T)}{2 \mathrm{i}\pi T}\right]$ is the mass enhancement factor due to the FLEX self-energy $\Sigma(k)$.
We obtain $Z(\bm{k},T)\simeq4.5$ for $\bm{k}=(\pi,0)$ at $T=0.01$. Therefore, we obtain the relation $\Delta^*_{\mathrm{BO}}(T=0)/T_{\mathrm{BO}}\simeq1.7$ for $\delta t_0=0.075$ and $T_\mathrm{BO}=0.02$ under the single-$\bm{q}$ BO.
It is almost the same as the BCS universal ratio. 
Note that $\Delta^*_{\mathrm{BO}}(T=0)/T_{\mathrm{BO}}\simeq3$ is observed in Hg-based and Y-based cuprates by the Raman spectroscopy \cite{gap-size}.
   
      Figures \ref{fig:DOS,Fermi}(c) and (d) show the FS with the unfolded spectral function under the single-$\bm{q}$ and the double-$\bm{q}$ BOs, respectively. The unfolded spectral function is obtained by $\rho(\bm{k},\omega)=-\frac{1}{\pi}\frac{1}{N_{c}}\sum_{\alpha \beta} \mathrm{Im}G^0_{\alpha \beta}(\bm{k},\omega)
 \mathrm{e}^{-\mathrm{i}\bm{k} \cdot (\bm{r}_\alpha-\bm{r}_\beta)}$ according to  Eq. (\ref{equ:unfold}),
 where $G^0_{\alpha \beta}(\bm{k},\omega)$ is the retarded bare Green function.  
 
The band-hybridization gap induces the Fermi arc structure. The FS near $\bm{k}=(0,\pm \pi)$ vanishes {and has $C_2$ symmetry under the single-$\bm{q}$ BO as shown in Fig. \ref{fig:DOS,Fermi} (d), so the electronic nematic response should appear}. On the other hand, the FS near $\bm{k}=(0,\pm \pi),(\pm \pi,0)$ vanishes  {and has the $C_4$ symmetry under the double-$\bm{q}$ BO shown in Fig. \ref{fig:intro} (d)  as shown in Fig. \ref{fig:DOS,Fermi} (d).  Interestingly, the FS under the double-$\bm{q}$ BO possesses the $C_4$ symmetry for any phase parameters $\gamma$ of two BOs, and therefore electronic nematic response is unexpected.}
The Fermi arc structure in Fig. \ref{fig:DOS,Fermi}(d) was actually observed by ARPES measurements \cite{T.Yoshida-Cu-ARPES,H.M.Fretwell,M.Vishik,shortrange}.
Even in the single-$\bm{q}$ case, if the BO has small domain structures, the Fermi arc structure similar to Fig. \ref{fig:DOS,Fermi}(d) would be also observed since the FS shown in Fig. \ref{fig:DOS,Fermi}(c) and its 90 degree rotation are averaged.   
Observed domain size is typically a few nm \cite{K.Fujita-Cu-STM,T.Hanagri,Y.Kohsaka,M.J.Lawler,W.Tabis2017}.
   Therefore, the experimental Fermi arc can be explained by both single-$\bm{q}$ and double-$\bm{q}$ BOs.
\begin{figure}[!htb]
\includegraphics[width=0.9\linewidth]{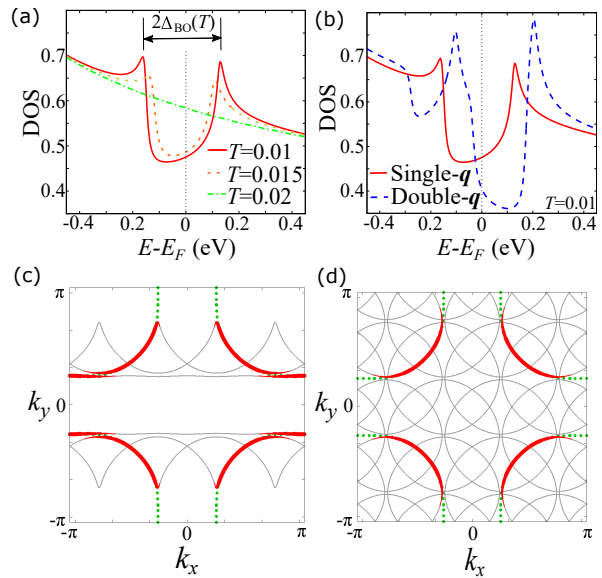}
\centering
\caption{
(a) $T$ dependence of the DOS under the single-$\bm{q}$ BO.
(b) DOS under the single-$\bm{q}$ and the double-$\bm{q}$ BOs at $T=0.01$.  
(c) Unfolded FS under the single-$\bm{q}$ BO (thick red lines) and the folded FS (thin gray lines) under the bond order. The green dotted lines represent
 the vanished FS  due to the BO.  (d) Folded and unfolded FSs under the double-$\bm{q}$ BO.
}
\label{fig:DOS,Fermi}
\end{figure}

\begin{figure}[!htb]
\includegraphics[width=.9\linewidth]{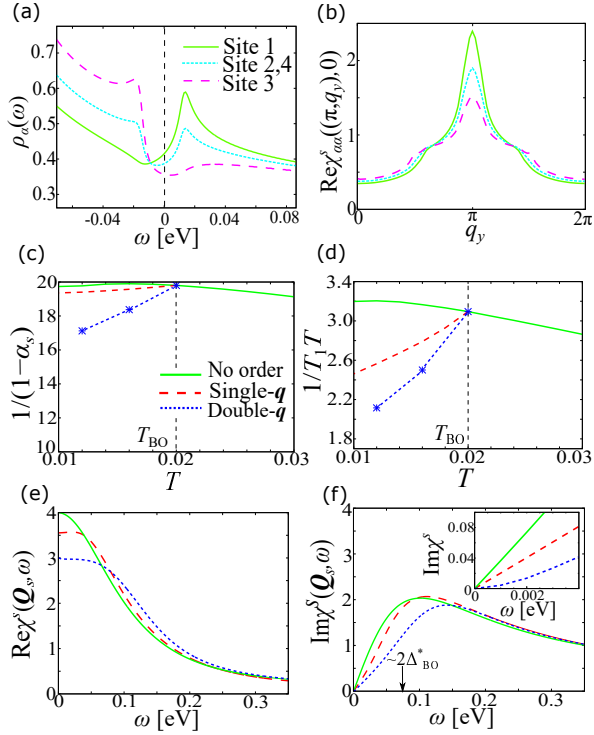}
\centering
\caption{
(a) $\rho_{\alpha}(\omega)$ and (b) $\chi_{\alpha\alpha}^s((\pi,q_y),0)$.
 The site numbers  are defined in Fig. \ref{fig:intro}(c). 
 (c) $T$ dependences of $1/(1-\alpha_s)$ and (d) that of $1/T_1T$. The solid, dashed, and dotted lines are no-order, single-$\bm{q}$, and double-$\bm{q}$ BOs, respectively.
(e) Re$\chi^s(\bm{Q}_s,\omega)$ and (f) Im$\chi^s(\bm{Q}_s,\omega)$ as a function of $\omega$ at $T=0.01$.  The inset shows Im$\chi^s(\bm{Q}_s,\omega)$ for the small 
$\omega$ region.
}
\label{fig:T-T1T}
\end{figure}

Next, we study $\hat{\chi}^s{(q)}$ with the four-period $d$-wave BO. 
Figures \ref{fig:T-T1T}(a) and (b) show the local DOS $\rho_{\alpha}(\omega)=-\frac{1}{\pi} \sum_{\bm{k}}\mathrm{Im}G^0_{\alpha \alpha}(\bm{k},\omega)$ and   
$\mathrm{Re}{\chi}^s_{\alpha\alpha}(\bm{q},0)$ for $\bm{q}=(\pi,q_y)$, respectively. $\alpha$ represents the site-numbers in Fig. \ref{fig:intro}(c).  For $\gamma=0$ in Eq. (\ref{equ:BO}), the components of the $\alpha=2$ and $\alpha=4$ are equivalent, and there are three inequivalent components.
We also show that there are two inequivalent components for $\gamma=\pi/4$ in Fig. \ref{fig:sp1}(b) and (c) in SM. Such local site-dependence has been observed as the broadening of the NMR signal \cite{site-dependence}.
 
 Figure \ref{fig:T-T1T}(c) shows the $T$ dependence of $1/(1-\alpha_s)$,  where $\alpha_s$ is the maximum eigenvalue of $U\hat{\chi}^0(\bm{q},0)$, and a magnetic transition emerges when $\alpha_s$ reaches unity because  $\hat{\chi}^s(\bm{Q}_s,0)$ diverges. 
 Figure \ref{fig:T-T1T}(d) shows the $T$ dependence of $1/T_1T$     
    given by
\begin{align}
1/T_1T \propto \frac{1}{N_c}\sum_{\alpha \beta}\sum_{\bm{q}}\lim_{\omega\rightarrow0}\frac{\mathrm{Im}\chi^s_{\alpha\beta}(\bm{q},\omega)}{\omega} \mathrm{e}^{-\mathrm{i}\bm{q} \cdot (\bm{r}_\alpha-\bm{r}_\beta)}.
\label{equ:1/T_1T}
\end{align} 
Both $1/T_1T$ and $\alpha_s$ decrease with decreasing $T$ below  $T_{\mathrm{BO}}$ due to the Fermi arc structure. 
The decrease in $\alpha_s$ due to the single-$\bm{q}$ BO is tiny maybe because the FS connected by the nesting vector $\bm{Q}_s$ in Fig. \ref{fig:intro}(b) remains  below $T_\mathrm{BO}$.
In a two-dimensional Fermi liquid, the relation between $1/T_1T$ and $\alpha_s$ is given by  
$1/T_1T \propto N^2(0)\frac{1}{1-\alpha_s}$            
for $T\simeq0$ \cite{Shiba, S-Moriya,K.Yamada}, where $N(0)$ is the DOS at the Fermi energy $E_F$. This equation indicates that $1/T_1T$ decreases in proportion to $N(0)^2$ even if the change of $\alpha_s$  is small.

In Fig. \ref{fig:T-T1T}(d), theoretical $1/T_1T$ shows a kink-type pseudogap, whereas experimental $1/T_1T$ exhibits a broad peak \cite{M.Takigawa-Cu-T1T}. 
This fact may indicate the existence of inhomogeneity in the BO state due to its short correlation length \cite{W.Tabis2017,dynamical,S.Blanco}.
Note that the kink-behavior in Fig. \ref{fig:T-T1T}(d) becomes moderate when $\Delta_\mathrm{BO}^*(T=0)/T_\mathrm{BO}$ is smaller or when $\alpha_s$ is larger at $T=T_\mathrm{BO}$.
We have also verified that the present numerical results for the four-period BO are essentially similar to the results in the five-period BO case.

    We discuss the relation between $1/T_1T$ and $\alpha_s$ in more detail.
     Figures \ref{fig:T-T1T}(e) and (f) show the real and imaginary parts of unfolded $\chi^s(\bm{Q}_s,\omega)$ for at $T=0.01$, respectively.
     $1/(1-\alpha_s)$  is proportional to $\chi^s(\bm{Q}_s,0)$, while $1/T_1T$ reflects the slope of Im$ \chi^s(\bm{Q}_s,\omega)$ for $\omega\simeq0$.
     As shown in Fig. \ref{fig:T-T1T}(f), Im$\chi^s(\bm{Q}_s,\omega)$ is suppressed for $\omega\lesssim2\Delta^*_\mathrm{BO}$, and its slope for small $\omega$ decreases.     
$\mathrm{Im}\chi^s(\bm{q},\omega)\simeq\frac{\mathrm{Im}\chi^0(\bm{q},\omega)}{(1-U\chi^0(\bm{q},0))^2}$ is satisfied
for small $\omega$, where $\chi^0(\bm{q},\omega)$ is the unfolded irreducible susceptibility.
It is known that $\mathrm{Im}\chi^0(\bm{q},\omega)$ is suppressed by a gap $\Delta^*_{\mathrm{BO}}$ for $\omega\lesssim 2\Delta^*_{\mathrm{BO}}$.  
 Re$\chi^s(\bm{q},\omega)$ and Im$\chi^s(\bm{q},\omega)$ are combined through the Kramers-Kroning relation 
$\mathrm{Re}\chi^s (\bm{q},\omega)=\frac{1}{\pi}\mathrm{P}\int_{-\infty} ^{\infty}\frac {\mathrm{Im}\chi^s(\bm{q},\omega')}{\omega'-\omega}
\mathrm{d}\omega'$. Since Re$\chi^s(\bm{q},\omega)$ is obtained by  $\omega$  integral of Im$\chi^s(\bm{q},\omega)$, 
the low-energy suppression of Im$\chi^s(\bm{Q}_s,\omega)$
 by the BO does not reduce Re$\chi^s(\bm{q},\omega)$ largely, except for the close vicinity of the magnetic quantum critical point.
 Therefore, the change of $\alpha_s$ is small irrespectively of the large decrease in $1/T_1T$.  

\begin{figure}[!htb]
\includegraphics[width=.9\linewidth]{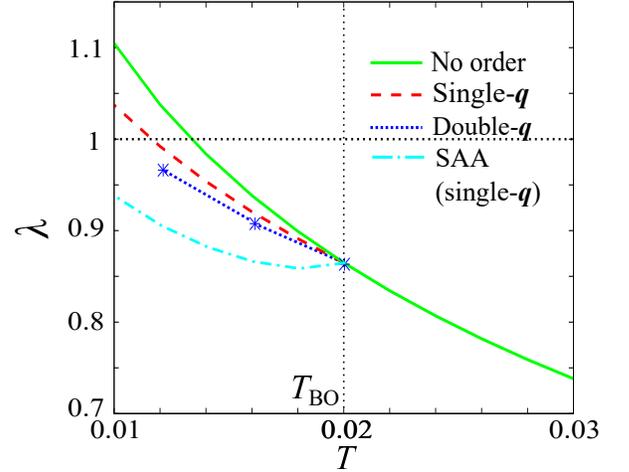}
\caption{
$T$ dependence of the eigenvalue $\lambda$ of the linearized Eliashberg equation. The solid, dashed, and dotted  lines correspond to no-order, single-$\bm{q}$ BO, and double-$\bm{q}$ BO, respectively. The dashed-dotted line represents $\lambda$ given by the SAA for the single-$\bm{q}$ BO. 
}
\label{fig:T-gap}
\end{figure}

Finally, we study the superconducting state by solving the linearized Eliashberg equation defined by
\begin{align}
 \lambda\Delta_{\alpha \beta}(k)=&\nonumber 
-\frac{T}{N}\sum_{k'\alpha' \beta'}
G_{\alpha\alpha'}(k')\Delta_{\alpha' \beta'}(k')G_{\beta\beta'}(-k')\\&\times V_{\alpha \beta}^{\mathrm{sc}}(k-k'),
\label{equ:gap}
\end{align}
where $V_{\alpha \beta}^{\mathrm{sc}}(q)=\frac
{3}{2}U^2\chi_{\alpha \beta}^s(q)-\frac{1}{2}
U^2\chi_{\alpha \beta}^c(q)+U\delta_{\alpha \beta}$
 is the singlet pairing interaction  \cite{D.J.Scalapino1986,D.J.Scalapino}.  
 Here, $\lambda$ is the eigenvalue, and $\Delta_{\alpha \beta}(k)$
 is the superconducting gap function. $N$ is the total number of the unit cells.
 At $T=T_c$, $\lambda=1$ is satisfied.
Figure \ref{fig:T-gap} shows the $T$ dependence of $\lambda$. 
We obtain $T_c=0.0134\simeq80$ K without the BO and $T_c=0.0116$ in the single-$\bm{q}$ BO.
 It means that $T_c$ is suppressed  only slightly (of order 10 K) due to the Fermi arc structure caused by the antiferro $d$-wave BO.
The suppression of $T_c$ by double-$\bm{q}$ BO is about twice of that by the single-$\bm{q}$ BO.   
These suppressions of $T_c$ are consistent with the experimental phase diagrams. \cite{W.Tabis,M.H,S.Blanco,suppressTc,3DCDWsuppress}. 
	 
	 In the present study, the site dependent electronic states under the BO are seriously analyzed. To verify that the site dependent analyses are important,  we perform another calculation method, the SAA,  in which calculations are performed based on the single-site model by using the unfolded bare Green function $G^0(k)$. 
 The eigenvalue $\lambda$ obtained by the SAA under the single-$\bm{q}$ BO  is significantly underestimated as explained in Fig. \ref{fig:T-gap}.
Also, both $1/T_1T$ and $1/(1-\alpha_s)$ are underestimated in the SAA as shown in Fig. \ref{fig:sp2}(a) and (b).
It is because the site dependence of spin susceptibility shown in Fig. \ref{fig:T-T1T}(b) is averaged in the SAA.
 Therefore, it is important to calculate the cluster-Hubbard model as performed in the present study.  

In summary, 
we studied the many-body electronic states to understand the CDW phase in Fig. \ref{fig:intro}(a). Based on the cluster Hubbard model with the finite $d$-wave BO, we analyzed $\hat{\chi}^s(\bm{q},\omega)$, $1/T_1T$, and $T_c$ by using the site-dependent FLEX approximation. 
We found that the strong suppression of $1/T_1T$ is induced by the BO while the decrease of $\alpha_s$ is tiny, consistently with experiments  \cite{M.Takigawa-Cu-T1T,neutron}.
We also found a slight decrease of $d$-wave $T_c$ about 10K under the  BO, which is also consistent with the experiments \cite{E.H.Shilva,W.Tabis,M.H,S.Blanco,suppressTc}.
Since $T_c$ and spin fluctuations in the antiferro BO are significantly underestimated by using the SAA, site dependence must be taken into account seriously.
This study supports the axial $d$-wave bond BO scenario in cuprates, where $T_{\mathrm{BO}}$ corresponds to $T_{\mathrm{CDW}}$ in Fig. \ref{fig:intro}(a).
 
   This work was supported by the ``Quantum Liquid Crystals'' No. JP19H05825 KAKENHI on Innovative Areas from JPSJ of Japan, and JSPS KAKENHI (JP18H01175, JP20K03858, JP17K05543).
 

\clearpage

\makeatletter
\renewcommand{\thefigure}{S\arabic{figure}}
\renewcommand{\theequation}{S\arabic{equation}}
\makeatother
\setcounter{figure}{0}
\setcounter{equation}{0}
\setcounter{page}{1}
\setcounter{section}{1}

\begin{widetext}
\begin{center}
{\bf 
[Supplementary Material] \\
Development of spin fluctuations 
under the presence of $d$-wave bond order in 
cuprate superconductors}%
\end{center}

\begin{center}
Satoshi Ando, Youichi Yamakawa, Seiichiro Onari, and Hiroshi Kontani
\end{center}

\begin{center}
\textit{Department of Physics, Nagoya University, Nagoya 464-8602, Japan}
\end{center}
\end{widetext}
\subsection{A: The site-dependent FLEX approximation in cluster Hubbard models}
We use the site-dependent FLEX approximation \cite{FLEX} to analyze $\hat{\chi}^s(q)$ and $T_c$ with the cluster Hubbard model. 
       The FLEX self-energy $\Sigma_{\alpha \beta}(k)$ is given  by 
\begin{align}
\Sigma_{\alpha \beta}(k)=\frac{T}{N}\sum_{{k'}} V_{\alpha \beta}(k-k')
G_{\alpha \beta}(k')
\label{equ:self-energy}
\end{align}
\begin{align}
V_{ \alpha \beta}(q)=\frac
{3}{2}U^2\chi_{\alpha \beta}^s(q)+\frac{1}{2}
U^2\chi_{\alpha \beta}^c(q).
\label{equ:self-potential}
\end{align} 
We use the short notations $k=(\bm{k},\mathrm{i}\varepsilon_n)$ and $q=(\bm{q},\mathrm{i}\omega_l)$, where $\varepsilon_n$ and $\omega_l$ are fermion and boson Matsubara frequencies, respectively. 
$V_{\alpha \beta}(k)$ is the effective interaction in the FLEX approximation, and  $G_{\alpha \beta}(k)$ is the Green function defined by
\begin{align}
G_{\alpha\beta}(\bm{k},\mathrm{i}\varepsilon_n)=\left[\mathrm{i}\varepsilon_n \hat{1}-\hat{H}(\bm{k})-\hat{\Sigma}(\bm{k},\mathrm{i}\varepsilon_n)+\mu\hat{1}\right]^{-1}_{\alpha\beta},
\label{equ:self-green}
\end{align} 
 where the components of $\hat{H}(\bm{k})$ is given by $H_{\alpha\beta}(\bm{k})=N^{-1}\sum_{lm}t_{lm,\alpha\beta}\exp{(-\mathrm{i}\bm{k}\cdot(\bm{R}_l-\bm{R}_m}))$. Here, $\bm{R}_l (\bm{R}_m)$ is the position of the unit-cell $l (m)$.
$\chi_{\alpha \beta}^c(q)$ is the charge susceptibility defined by ${\chi}^c_{\alpha\beta}(q)=\left[\hat{\chi}^0(q)(1+U\hat{\chi}^0(q))^{-1}\right]_{\alpha\beta}$, and ${\chi}^s_{\alpha\beta}(q)$ is the spin susceptibility defined by ${\chi}^s_{\alpha\beta}(q)=\left[\hat{\chi}^0(q)(1-U\hat{\chi}^0(q))^{-1}\right]_{\alpha\beta}$,
where the irreducible susceptibility ${\chi}^0_{\alpha\beta}(q)$ is defined by $\chi_{\alpha\beta}^0({q})=-{T}{N}^{-1}\sum_{{k}} G_{\alpha\beta}({k}+{q})G_{\beta\alpha}({k})$. 
$N$  is the total number of the unit cells, and $\mu$ is the chemical potential.
  We solve  these Eqs. (\ref{equ:self-energy})-(\ref{equ:self-green}) self-consistently.
In the present study, we subtract the static and Hermite part of the self-energy ``$\Delta\Sigma_{\alpha \beta}(\bm{k})$'' from $\Sigma_{\alpha \beta}(k)$ to keep the shape of the Fermi surface. 
$\Delta\Sigma_{\alpha \beta}(\bm{k})$ is given by    
$\Delta \Sigma_{\alpha \beta}(\bm{k})\equiv \frac{1}{2}\left[\Sigma_{\alpha \beta}(\bm{k},0_+)+\Sigma_{\alpha \beta}(\bm{k},0_-)\right]$, where 
$\Sigma_{\alpha \beta}(\bm{k},0_{\pm})\sim \frac{3}{2}\Sigma_{\alpha\beta}(\bm{k},\pm  
\mathrm{i} \pi T)-\frac{1}{2}\Sigma_{\alpha\beta}(\bm{k},\pm \mathrm{i}3 \pi T)$.   

\subsection{B: Independence of the phase of the anti ferro $d$-wave bond order }

\begin{figure}[!htb]
\includegraphics[width=.9\linewidth]{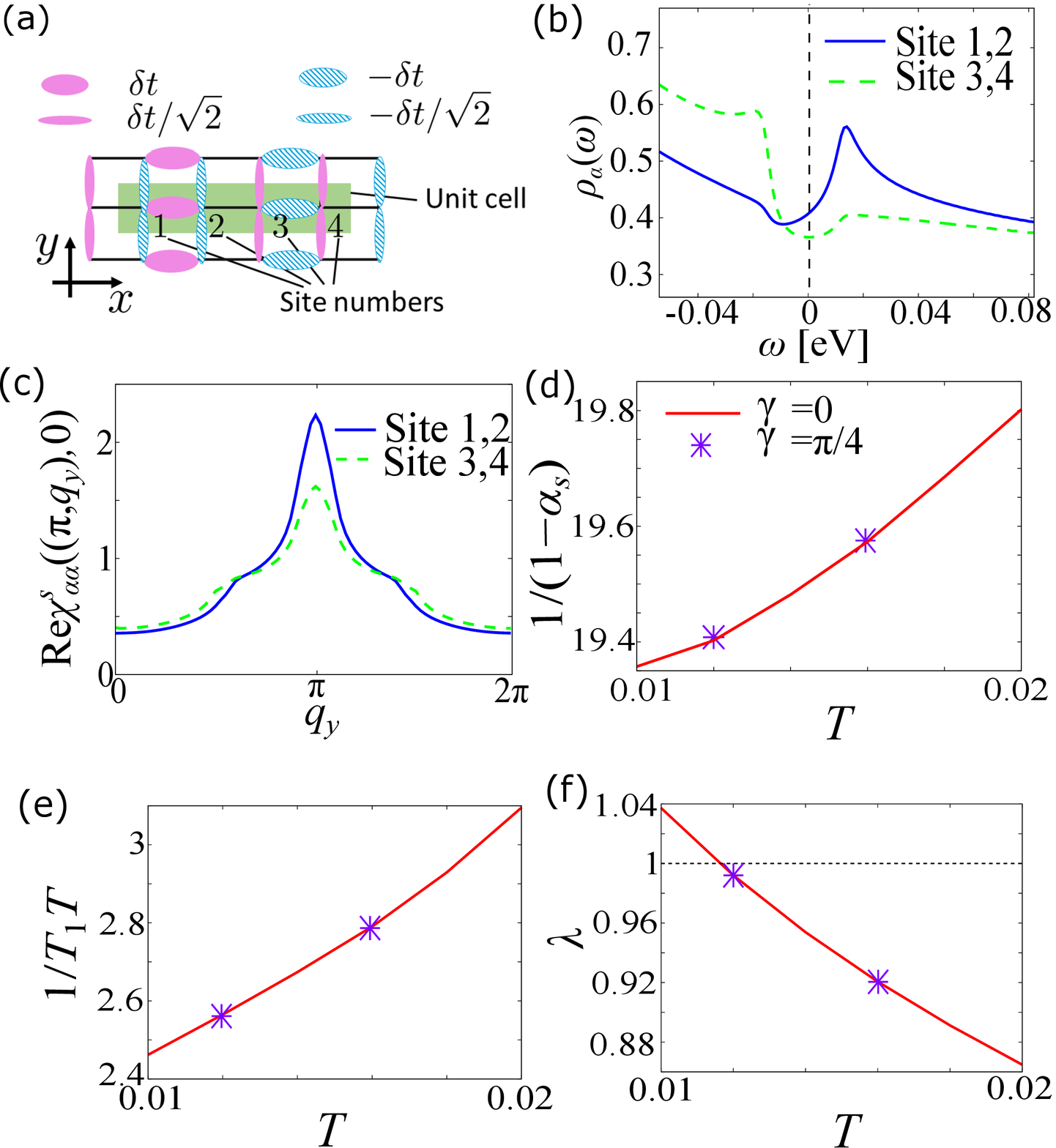}
\caption{
(a) Schematic single-$\bm{q}$ four-period $d$-wave BO for $\gamma=\pi/4$. $\delta t$ is the modulation of the hopping.  
(b) $\rho_{\alpha}(\omega)$ and (c) Re$\chi_{\alpha\alpha}^s((\pi,q_y),0)$ for $\gamma=\pi/4$ at $T=0.01$
(d) $T$ dependence of $1/(1-\alpha_s)$, (e)  that of $1/T_1T$, and (f) that of $\lambda$ for $\gamma=0$ and $\gamma=\pi/4$ under the single-$\bm{q}$ BO.  
}
\label{fig:sp1}
\end{figure}
In the main text, we set  $\gamma=0$ as the phase of the bond order (BO) in Eq. (\ref{equ:BO}). However, the effects of $\gamma$ on the physical quantities are not trivial. 
Here, we also investigate the $\gamma=\pi/4$ case.
Figure \ref{fig:sp1}(a) is the schematic picture of the four-period $d$-wave BO for $\gamma=\pi/4$.
 Then, the local DOS $\rho_{\alpha}(\omega)$ and $\chi^s_{\alpha\alpha}((\pi,q_y),0)$ have the site dependence and split into two lines as shown in Figs. \ref{fig:sp1}(b) and (c),
 which are different from the results for $\gamma=0$ in the main text in Figs. \ref{fig:T-T1T}(a) and (b).
On the other hand, the total DOS in Fig. \ref{fig:DOS,Fermi}(a) for $\gamma=0$ is equivalent to that for $\gamma=\pi/4$ completely.
 As shown in Fig. \ref{fig:sp1}(d), $\alpha_s$ also does not depend on $\gamma$ in spite of 
 the difference between $\chi^s_{\alpha\alpha}$ for $\gamma=0$ in Fig. \ref{fig:T-T1T}(c) and that for $\gamma=\pi/4$.      
Also, $1/T_1T$ and $\lambda$ are independent on $\gamma$ as shown in Figs. \ref{fig:sp1}(e) and (f).

In the $4\times1$ model, 
the dominant inter-band hybridization on the FS can be understood by a $2\times2$ matrix.
At $\bm{k}=(0,\pm\pi)$,
 ${\bm{k}\pm{\bm{Q}_1}/{2}}$ are on the FS as shown in  Fig. \ref{fig:intro} (b), and the band dispersions without order satisfy the relation $\varepsilon_{\bm{k}\pm{\bm{Q}_1}/{2}}\approx \mu$. 
The hybridization between ${\bm{k}-{\bm{Q}_1}/{2}}$ and ${\bm{k}+{\bm{Q}_1}/{2}}$ is represented by
\[
   \left(
    \begin{array}{cc}
      \varepsilon_{\bm{k}-{\bm{Q}_1}/{2}}-\mu & \delta t_{\bm{k}-{\bm{Q}_1}/{2},\bm{k}+{\bm{Q}_1}/{2}}  \\
        \delta t^*_{\bm{k}-{\bm{Q}_1}/{2},\bm{k}+{\bm{Q}_1}/{2}}  & \varepsilon_{\bm{k}+{\bm{Q}_1}/{2}}-\mu 
    \end{array}
  \right),
\]
  where $\delta t_{\bm{k}-{\bm{Q}_1}/{2},\bm{k}+{\bm{Q}_1}/{2}}$ is the band hybridization component due to the BO and is proportional to ``$d$-wave form factor'', which is defied by   
\begin{align}
f_d(\bm{k})\equiv\cos{k_x}-\cos{k_y}.
\label{equ:dwave}
\end{align} 
 $\delta t_{\bm{k}-{\bm{Q}_1}/{2},\bm{k}+{\bm{Q}_1}/{2}}$ is given by  
 $\delta t_{\bm{k}-{\bm{Q}_1}/{2},\bm{k}+{\bm{Q}_1}/{2}}
 =\\N_{c}^{-1}\sum_{ij} \delta t_{ij}^{\bm{Q}_1}
\exp{\left[\mathrm{i}(\bm{k}-{\bm{Q}_1}/{2})\cdot\bm{r}_i\right]}\exp{\left[-\mathrm{i}(\bm{k}+{\bm{Q}_1}/{2})\cdot\bm{r}_j\right]}\\=\delta t f_d(\bm{k})$. 
  Then, the pseudogap opens due to $|\delta t_{\bm{k}-{\bm{Q}_1}/{2},\bm{k}+{\bm{Q}_1}/{2}}|=2\delta t(T)$ at $(0,\pm\pi)$, and the gap size is $2\Delta_\mathrm{BO}(T)\simeq4\delta t (T)$ as shown in Fig. \ref{fig:DOS,Fermi} (a).
     $|\delta t_{\bm{k}-{\bm{Q}_1}/{2},\bm{k}+{\bm{Q}_1}/{2}}|$ does not depend on $\gamma$, therefore the total DOS is also independent on $\gamma$.   

\subsection{C: Site-averaging approximation (SAA)  results}
\begin{figure}[!htb]
\includegraphics[width=0.9\linewidth]{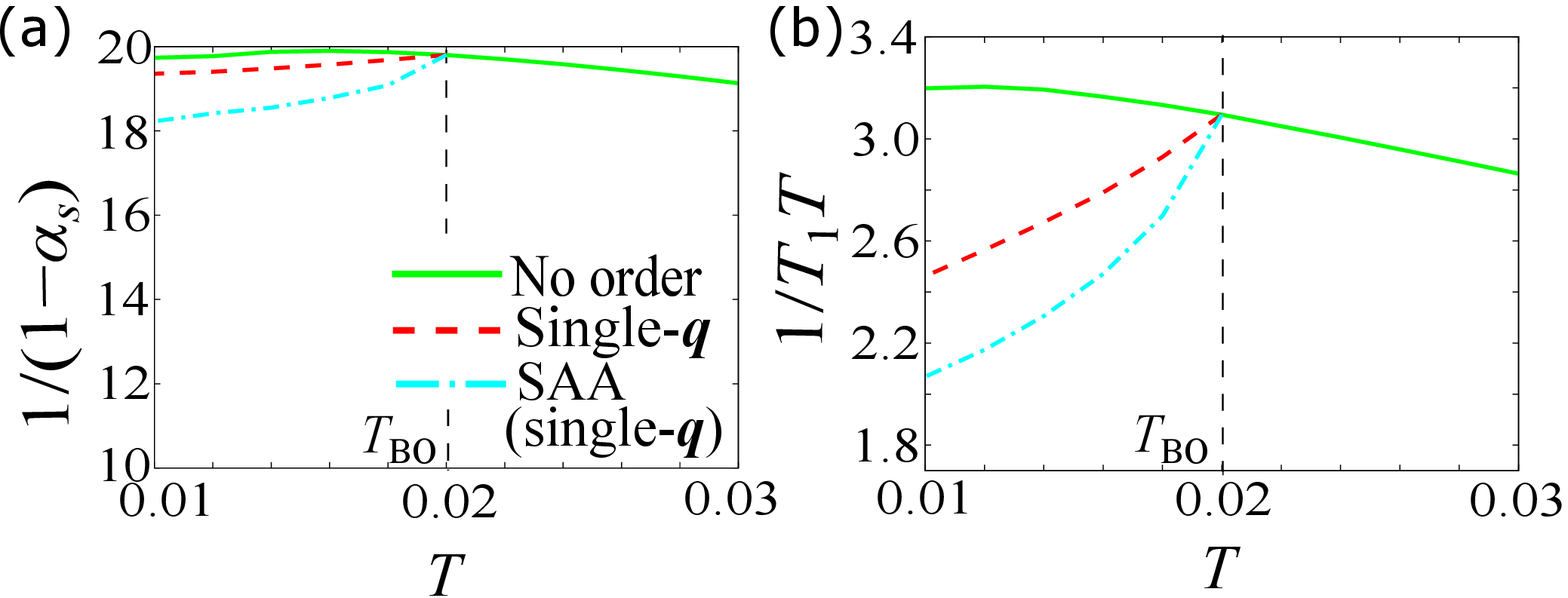}
\centering
\caption{
(a) $T$ dependence of $1/(1-\alpha_s)$ and (b) that of $1/T_1T$ obtained by the single-$\bm{q}$ BO and the SAA for the single-$\bm{q}$ BO.   
}
\label{fig:sp2}
\end{figure}

In the site-averaging approximation (SAA) introduced in the main text, we use the unfolded bare Green function $G^0({k})$, 
which is the site-average of $G^0_{\alpha\beta}({k})=\left[(\mathrm{i}\varepsilon_n \hat{1}-\hat{H}(\bm{k})+\mu\hat{1})^{-1}\right]_{\alpha\beta}$ following Eq. (\ref{equ:unfold}). 
$G^0({k})$ reproduces the total DOS shown in Fig. \ref{fig:DOS,Fermi} (a) under the antiferro BOs, while the site-dependence is dropped.
 The SAA Green function is defined by
 \begin{align}
 G_{\mathrm{SAA}}(k)=G^0(k)/(1-G^0(k)\Sigma_{\mathrm{SAA}}(k)),
 \label{SAA}
 \end{align} 
 where $\Sigma_{\mathrm{SAA}}(k)$ is the single-site FLEX self-energy composed of
 $G_{\mathrm{SAA}}(k)$.
  We solve $G_{\mathrm{SAA}}(k)$ in Eq. (\ref{SAA}) and  $\Sigma_{\mathrm{SAA}}(k)$ self-consistently in the SAA-FLEX.    
  In Fig.  \ref{fig:sp2}, we compare the results of the cluster model with those in the SAA for single-$\bm{q}$ BO.
  Figures \ref{fig:sp2}(a) and (b) exhibit the obtained  $1/(1-\alpha_s)$ and $1/T_1T$, respectively. 
  Both of them are underestimated in the SAA.
  In the same way, $\lambda$ given by the SAA in Fig. \ref{fig:T-gap} is prominently underestimated. 
  Those underestimation come from the neglect of  the site dependence in $\chi^s_{\alpha\beta}$ in Fig. \ref{fig:T-T1T}(b) or \ref{fig:sp1}(c).

\subsection{D: DOS under the $s'$-wave BO}

\begin{figure}[!htb]
\vspace*{1cm}
\includegraphics[width=0.9\linewidth]{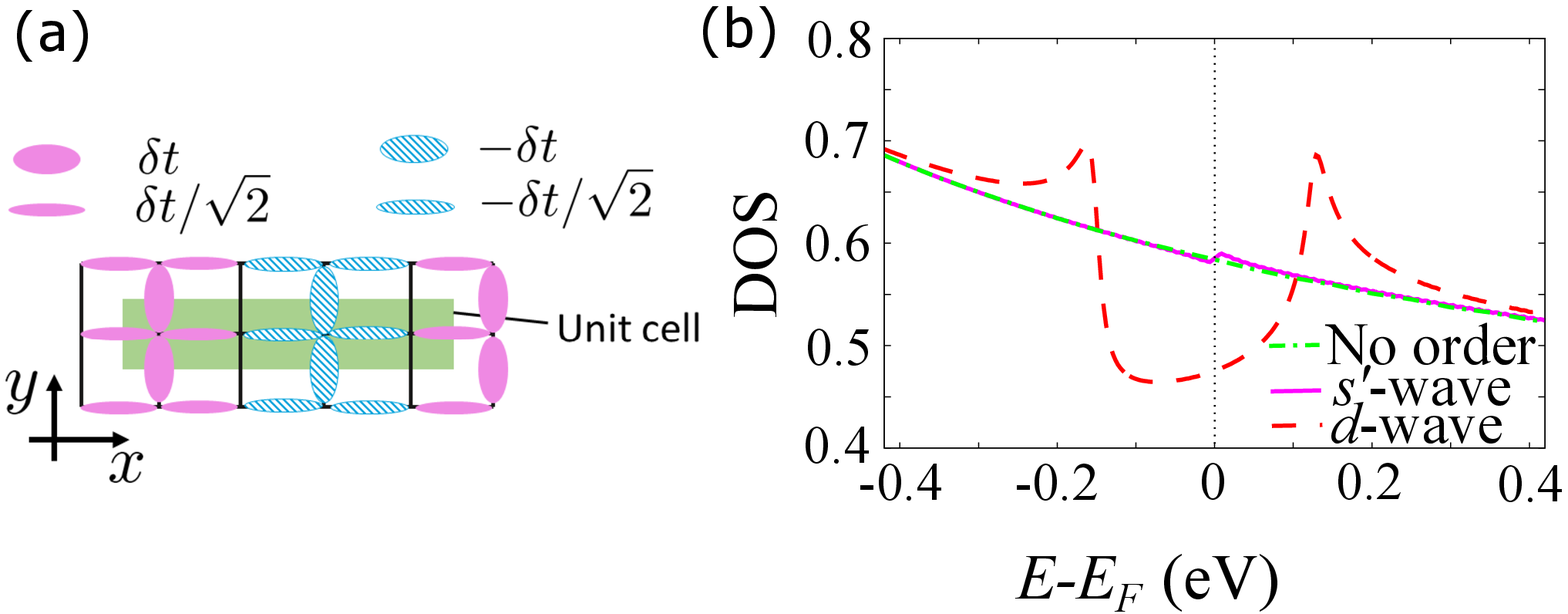}
\centering
\caption{
(a) Schematic single-$\bm{q}$ four-period $s'$-wave BO. $\delta t$ is the modulation of the hopping.  
(b) DOS under the single-$\bm{q}$ $s'$-wave BO, the single-$\bm{q}$  $d$-wave BO, and no-order at $T=0.01$.     
}
\label{fig:sp4}
\end{figure}
We consider the two types $s$-wave form factors ``$f_{s'}(\bm{k})\equiv\cos{k_x}+\cos{k_y}$'' and ''$f_s(\bm{k})\equiv1$''. 
 The Fourier-transformation of them gives the $s'$-wave BO $\delta t^{\bm{q}}_{ij} \propto\left[\delta_{\bm{r}_i-\bm{r}_j,\hat{\bm{x}}}+\delta_{\bm{r}_i-\bm{r}_j,\hat{\bm{y}}}+(i\leftrightarrow j)\right]\mathrm{Re}\{\mathrm{e}^{-\mathrm{i}\left(\bm{q}\cdot\frac{\bm{r}_i+\bm{r}_j}{2} \right)}\}$ as shown in Fig. \ref{fig:sp4} (a) and the conventional CDW $\delta t^{\bm{q}}_{ij} \propto \delta_{\bm{r}_i,\bm{r}_j}\mathrm{Re}\{\mathrm{e}^{-\mathrm{i}\left(\bm{q}\cdot\frac{\bm{r}_i+\bm{r}_j}{2} \right)}\}$, respectively.
 In the $s'$-wave BO case, the pseudogap is not induced as shown in  Fig. \ref{fig:sp4} (b) because $f_{s'}(\bm{k})\propto \delta t_{\bm{k}-{\bm{Q}_1}/{2},\bm{k}+{\bm{Q}_1}/{2}}$ is zero at ($0,\pm \pi$) and the relation $\varepsilon_{\bm{k}\pm{\bm{Q}_1}/{2}}\approx \mu$ holds.   
  The $s'$-wave BO is observed in La-based compound (LBCO) \cite{A.J.Achkar}. Our analysis for the $s'$-wave BO is consistent that apparent pseudogap have not been observed in LBCO.
  On the other hand,  the conventional CDW with the local density modulation induces the pseudogap because $f_{s}(\bm{k})$ has a finite value at ($0,\pm \pi$). However, it is strongly suppressed by the on-site Coulomb interaction $U$ and does not realize in cuprate superconductors.  
  




\begin{thebibliography}{99}
\bibitem{D.Senechal}
D. Senechal and A.-M.S. Tremblay, Phys. Rev. Lett. {\bf 92},
126401 (2004).
\bibitem{B. Kyung}
B. Kyung, S. S. Kancharla, D. Senechal, A. -M. S. Tremblay,
M. Civelli, and G. Kotliar, Phys. Rev. B {\bf 73}, 165114 (2006).
\bibitem{T.A.Maier}
T.A. Maier, M.S. Jarrell, and D.J. Scalapino, Physica C,
{\bf 460-462}, 13 (2007).

\bibitem{A.Shekhter}
 A. Shekhter, B. J. Ramshaw, R. Liang, W. N. Hardy, D. A. Bonn, F. F.
Balakirev, R. D. McDonald, J. B. Betts, S. C. Riggs, and A. Migliori,
Nature 498, {\bf 75} (2013).
\bibitem{R.-H.He}
 R.-H. He, M. Hashimoto, H. Karapetyan, J. D. Koralek, J. P. Hinton,
J. P. Testaud, V. Nathan, Y. Yoshida, H. Yao, K. Tanaka, W.
Meevasana, R. G. Moore, D. H. Lu, S.-K. Mo, M. Ishikado, H. Eisaki,
Z. Hussain, T. P. Devereaux, S. A. Kivelson, J. Orenstein, A.
Kapitulnik, and Z.-X. Shen, Science {\bf 331}, 1579 (2011).
\bibitem{Y.Sato}
 Y. Sato S. Kasahara, H. Murayama, Y. Kasahara, E.-
G. Moon, T. Nishizaki, T. Loew, J. Porras, B. Keimer,
T. Shibauchi, and Y. Matsuda, Nat. Phys. {\bf 13}, 1074 (2017).

\bibitem{loopcurrent}
B. Fauqué, Y. Sidis, V. Hinkov, S. Pailhès, C. T. Lin, X. Chaud, and P. Bourges, 
Phys. Rev. Lett. {\bf 96}, 197001 (2006).

\bibitem{S-Tsuchiizu-Cu}   
M. Tsuchiizu, K. Kawaguchi, Y. Yamakawa, and H. Kontani, Phys. Rev. B
	{\bf 97}, 165131 (2018).
\bibitem{S-Kawaguchi-Cu}
K. Kawaguchi, M. Tsuchiizu, Y. Yamakawa, and H. Kontani, 
J. Phys. Soc. Jpn. {\bf 86}, 063707 (2017).	
\bibitem{Y.Yamakawa-Cu}
Y. Yamakawa and H. Kontani, Phys. Rev. Lett. {\bf114}, 257001 (2015).
	
\bibitem{spin-loop-Kontani}
H. Kontani, Y. Yamakawa, R. Tazai, and S. Onari,
Phys. Rev. Research {\bf 3}, 013127 (2021).
\bibitem{C.M.Varma}
C. M. Varma, Phys. Rev. B {\bf 55}, 14554 (1997).
\bibitem{I.Affleck}
I. Affleck and J. B. Marston, Phys. Rev. B {\bf 37}, 3774(R) (1988).
\bibitem{F.C.Zhang}
 F. C. Zhang, Phys. Rev. Lett. {\bf 64}, 974 (1990).
\bibitem{R.Tazai-chargeloop}
R. Tazai, Y. Yamakawa, and H. Kontani, arXiv:2010.16109. 



\bibitem{G.Ghiringhelli}
G. Ghiringhelli, M. L. Tacon, M. Minola, S. Blanco-Canosa, C.
Mazzoli, N. B. Brookes, G. M. D. Luca, A. Frano, D. G. Hawthorn, F.
He, T. Loew, M. M. Sala, D. C. Peets, M. Salluzzo, E. Schierle, R.
Sutarto, G. A. Sawatzky, E. Weschke, B. Keimer, and L. Braicovich,
Science {\bf 337}, 821 (2012).
\bibitem{J.Chang}
J. Chang, E. Blackburn, A. T. Holmes, N. B. Christensen, J. Larsen,
J. Mesot, R. Liang, D. A. Bonn, W. N. Hardy, A. Watenphul, M. v.
Zimmermann, E. M. Forgan, and S. M. Hayden, Nat. Phys. {\bf 8}, 871
(2012).
\bibitem{E.Blackburn}
E. Blackburn, J. Chang, M. Hücker, A. T. Holmes, N. B. Christensen,
R. Liang, D. A. Bonn, W. N. Hardy, U. Rütt, O. Gutowski, M. v.
Zimmermann, E. M. Forgan, and S. M. Hayden, Phys. Rev. Lett. { \bf 110},
137004 (2013).
\bibitem{M.H2011}
5 M. Hücker, M. v. Zimmermann, G. D. Gu, Z. J. Xu, J. S. Wen, G. Xu,
H. J. Kang, A. Zheludev, and J. M. Tranquada, Phys. Rev. B {\bf 83},
104506 (2011).
\bibitem{dynamical}
F.Boschini, M.Minola R.Sutarto, E.Schierle, M.Bluschke, S.Das, Y.Yang, M.Michiardi, Y.C.Shao, X.Feng, S.Ono, R.D.Zhong, J.A.Schneeloch, G.D.Gu, E.Weschke, F.He, Y.DChuang, B.Keimer, A.Damascelli, A.Frano, and E.H.da Silva Neto, Nat. Commun. {\bf 12}, 597 (2021).
\bibitem{E.H.Shilva}
E. H. da Silva Neto, P. Aynajian, A. Frano, R. Comin, E. Schierle, E.
Weschke, A. Gyenis, J. Wen, J. Schneeloch, Z. Xu, S. Ono, G. Gu,
M. L. Tacon, and A. Yazdani, Science {\bf 343}, 393 (2014).
\bibitem{R.Comin}
R. Comin and A. Damascelli, Annu. Rev. Condens. Matter Phys. {\bf 7}, 369 (2016).
\bibitem{R.Comin2015}
R. Comin, R. Sutarto, F. He, E. da Silva Neto, L. Chauviere, A. Frano,
R. Liang, W. N. Hardy, D. Bonn, Y. Yoshida, H. Eisaki, J. E. Hoffman,
B. Keimer, G. A. Sawatzky, and A. Damascelli, Nat. Mater. {\bf 14}, 796
(2015).
\bibitem{R.Comin2014}
R. Comin, A. Frano, M. M. Yee, Y. Yoshida, H. Eisaki, E. Schierle, E.
Weschke, R. Sutarto, F. He, A. Soumyanarayanan, Y. He, M. L. Tacon,
I. S. Elfimov, J. E. Hoffman, G. A. Sawatzky, B. Keimer, and A.
Damascelli, Science {\bf 343}, 390 (2014).
\bibitem{W.Tabis2017}
W. Tabis, B. Yu, I. Bialo, M. Bluschke, T. Kolodziej,
A. Kozlowski, E. Blackburn, K. Sen, E. M. Forgan, M. v.
Zimmermann, Y. Tang, E.Weschke, B. Vignolle, M. Hepting,
H. Gretarsson, R. Sutarto, F. He, M. Le Tacon,
N. Bari\u{s}i´c, G. Yu, and M. Greven,
Phys. Rev. B 96, 134510 (2017).
\bibitem{fluctuation-Cu}
R. Arpaia, S. Caprara, R. Fumagalli, G. De Vecchi, Y. Y. Peng, E. Andersson,
D. Betto, G. M. De Luca, N. B. Brookes, F. Lombardi, M. Salluzzo, L. Braicovich,
C. Di Castro, M. Grilli, G. Ghiringhelli
,Science {\bf 365}, 6456 (2019). 


\bibitem{W.Tabis}
 W. Tabis, Y. Li, M. L. Tacon, L. Braicovich, A. Kreyssig, M. Minola,
G. Dellea, E. Weschke, M. J. Veit, M. Ramazanoglu, A. I. Goldman, T.
Schmitt, G. Ghiringhelli, N. Barisić, M. K. Chan, C. J. Dorow, G. Yu,
X. Zhao, B. Keimer, and M. Greven, Nat. Commun. {\bf 5}, 5875 (2014).
\bibitem{M.H}
M. Hücker, N. B. Christensen, A. T. Holmes, E. Blackburn, E. M. Forgan, Ruixing Liang, D. A. Bonn, W. N. Hardy, O. Gutowski, M. v. Zimmermann, S. M. Hayden, and J. Chang, Phys. Rev. B {\bf 90}, 054514 (2014).
\bibitem{S.Blanco}
S. Blanco-Canosa, A. Frano, E. Schierle, J. Porras, T. Loew, M. Minola, M. Bluschke, E. Weschke, B. Keimer, and M. Le Tacon,
Phys. Rev. B {\bf 90}, 054513 (2014).
\bibitem{suppressTc}
Naman K. Gupta, C. McMahon, R. Sutarto, T. Shi
, R. Gong, Haofei I. Wei, K. M.
Shen, F. He, Q. Ma, M. Dragomir, B. D. Gaulin, D. G. Hawthorn,
arXiv:2012.08450.


\bibitem{K.Fujita-Cu-STM}
 K. Fujita, M. Hamidian, S. Edkins, C. Kim, Y. Kohsaka, M. Azuma, M. Takano, H. Takagi, H. Eisaki, S. Uchida, A. Allais, M. J. Lawler, E. -A. Kim, S. Sachdev, and J. C. S. Davis, Proc. Natl. Acad. Sci. U.S.A.{ \bf 111}, E3026 (2014).
 \bibitem{T.Hanagri}
  T. Hanaguri, C. Lupien, Y. Kohsaka, D.-H. Lee, M. Azuma, M.
Takano, H. Takagi, and J. C. Davis, Nature {\bf430}, 1001 (2004).
\bibitem{Y.Kohsaka}
Y. Kohsaka, T. Hanaguri, M. Azuma, M. Takano, J. C. Davis, and H.
Takagi, Nat. Phys. {\bf 8}, 534 (2012).
\bibitem{M.J.Lawler}
M. J. Lawler, K. Fujita, J. Lee, A. R. Schmidt, Y. Kohsaka, C. K. Kim,
H. Eisaki, S. Uchida, J. C. Davis, J. P. Sethna, and E.-A. Kim, Nature
{\bf 466}, 347 (2010).

\bibitem{3DCDWsuppress}
M. E. Barber,  H. Kim, T. Loew, M. L. Tacon, M. Minola,
M. Konczykowski, B. Keimer, A. P. Mackenzie, and C. W. Hicks,
arXiv:2101.02923.

\bibitem{J.C.Davis}
 J. C. Davis and D.-H. Lee, Proc. Natl. Acad. Sci. USA
{\bf 110}, 17623 (2013).
\bibitem{M.A.Metlitski}
 M. A. Metlitski and S. Sachdev, Phys. Rev. B {\bf 82}, 075128 (2010).
\bibitem{C.Husemann}
 C. Husemann and W. Metzner, Phys. Rev. B {\bf 86}, 085113 (2012).
\bibitem{K.B.Efetov}
 K. B. Efetov, H. Meier, and C. P´epin, Nat. Phys. {\bf 9}, 442
(2013).
\bibitem{S.Sachdev}
 S. Sachdev and R. La Placa, Phys. Rev. Lett. {\bf 111}, 027202
(2013).
\bibitem{V.Mishra}
 V. Mishra and M. R. Norman, Phys. Rev. B {\bf 92}, 060507
(2015).	

\bibitem{E.Berg}
E. Berg, E. Fradkin, S. A. Kivelson, and J. M. Tranquada,
New J. Phys. {\bf 11}, 115004 (2009).
\bibitem{P.A.Lee}
P. A. Lee, Phys. Rev. X {\bf 4}, 031017 (2014).
\bibitem{E.Fradkin}
E. Fradkin, S. A. Kivelson, and J. M. Tranquada, Rev.
Mod. Phys. {\bf 87}, 457 (2015).
\bibitem{Y.Wang}
Y. Wang, D. F. Agterberg, and A. V. Chubukov, Phys.
Rev. Lett. {\bf 114}, 197001 (2015).

\bibitem{S-Onari-form}
S. Onari, Y. Yamakawa, and H. Kontani,
 Phys. Rev. Lett. {\bf 116}, 227001 (2016).
\bibitem{S.onari2012}
S. onari and H. Kontani,  Phys. Rev.
Lett. {\bf 109}, 137001 (2012).
\bibitem{Y.Yamakawa2016}
Y. Yamakawa, S. Onari, and H. Kontani,
Phys. Rev. X {\bf 6}, 021032 (2016).
\bibitem{R.Tazai2020}
R. Tazai, Y. Yamakawa, M. Tsuchiizu, and H. Kontani,
arXiv:2010.15516.
\bibitem{S.Onari2019}
S. Onari and H. Kontani,  Phys. Rev. B {\bf 100}, 020507(R) (2019).
\bibitem{S.Onari2020}
 S. Onari and H. Kontani, Phys. Rev. Res. {\bf 2}, 042005(R) (2020).
 
\bibitem{W.Metzner}	
W. Metzner, M. Salmhofer, C. Honerkamp, V. Meden,
and K. Sch¨onhammer, Rev. Mod. Phys. {\bf 84}, 299 (2012).
\bibitem{C.Bourbonnais}
C. Bourbonnais, B. Guay, and R. Wortis, in Theoret-
ical Methods for Strongly Correlated Electrons, edited
by D. S\'{e}n\'{e}chal, A.-M. Tremblay, and C. Bourbonnais
(Springer, New York, 2004) pp. 77-137.	
\bibitem{M.Tsuchiizu2013}
M. Tsuchiizu, Y. Ohno, S. Onari, and H. Kontani, Phys.
Rev. Lett. {\bf 111}, 057003 (2013).
\bibitem{R.Tazai2016}
R. Tazai, Y. Yamakawa, M. Tsuchiizu, and H. Kontani, Phys. Rev. B {\bf 94}, 115155 (2016).
\bibitem{C.Honerkamp}
 C. Honerkamp, ibid. {\bf 72}, 115103 (2005).
 
 \bibitem{Y.Wang2014}
Y. Wang and A. V. Chubukov, Phys. Rev. B {\bf 90}, 035149 (2014).
 \bibitem{tazai2019}
 R. tazai and H. Kontani, Phys. Rev. B {\bf 100}, 241103(R)
 
\bibitem{H.Kontani}
H. Kontani, Rep. Prog. Phys. {\bf 71} 026501 (2008).
\bibitem{H.Kontani-Fermi-liquid}
H. Kontani, Transport Phenomena in Strongly Correlated Fermi
Liquids (Springer-Verlag, Berlin, 2013).
 
\bibitem{hall-seebeck}
Nicolas Doiron-Leyraud, S. Lepault, O. Cyr-Choinière, B. Vignolle, G. Grissonnanche, F. Laliberté, J. Chang, N. Bari\u{s}i\'{c}, M. K. Chan, L. Ji, X. Zhao, Y. Li, M. Greven, C. Proust, and Louis Taillefer, 
Phys. Rev. X {\bf 3}, 021019 (2013).
\bibitem{hall}
David LeBoeuf, Nicolas Doiron-Leyraud, Julien Levallois, R. Daou, J.-B. Bonnemaison, N. E. Hussey, L. Balicas, B. J. Ramshaw, Ruixing Liang, D. A. Bonn, W. N. Hardy, S. Adachi, Cyril Proust, Louis Taillefer,
Nature {\bf 450}, 533 (2007).
\bibitem{seebeck}
F. Laliberté, J. Chang, N. Doiron-Leyraud, E. Hassinger , R. Daou, M. Rondeau , B.J. Ramshaw , R. Liang ,D.A. Bonn , W.N. Hardy , S. Pyon, T. Takayama4, H. Takagi
, I. Sheikin , L. Malone , C. Proust ,K. Behnia
 and Louis Taillefer,
Nat. Commun. {\bf 2}, 432 (2011).

\bibitem{M.Takigawa-Cu-T1T}
M. Takigawa, A. P. Reyes, P. C. Hammel, J. D. Thompson, R. H. Heffner, Z. Fisk, and
K. C. Ott, PRB. {\bf 43}, 247 (1991).


\bibitem{neutron}
M. K. Chan, C. J. Dorow, L. Mangin-Thro, Y. Tang, Y. Ge, M. J. Veit, G. Yu, X. Zhao, A. D. Christianson, J. T. Park, Y. Sidis, P. Steffens, D. L. Abernathy, P. Bourges and M. Greven, 
Nat. Commun. {\bf 7},  10819 (2016).

\bibitem{FLEX1}
N. E. Bickers and S. R. White, Phys. Rev. B {\bf 43}, 8044 (1991).

\bibitem{P.Hansmann}
P.Hansmann, N. Parragh, A. Toschi, G. Sangiovanni, and K. Held, New J. Phys. {\bf 16} 033009 (2014).

\bibitem{unfold}
W. Ku, T. Berlijn, and C, -C, Lee, Phys. Rev. Lett. {\bf104}, 216401 (2010).

\bibitem{reconst-FS}
K. Seo, and S. Tewari, Phys. Rev. B {\bf 90}, 174503 (2014). 

\bibitem{gap-size}
B. Loret, N. Auvray, Y. Gallais, M. Cazayous, A. Forget, D. Colson, M.-H. Julien, I. Paul, M. Civelli and A. Sacuto, 
Nat. Phys.  {\bf 15}, 771-775(2019)

\bibitem{T.Yoshida-Cu-ARPES}
T. Yoshida, M. Hashimoto, I. M. Vishik, Z. -X. Shen and A. Fujimori
, JPSJ. {\bf81}, 011006(2012).
\bibitem{H.M.Fretwell}
H. M. Fretwell, A. Kaminski, J. Mesot, J. C. Campuzano,1, M. R. Norman, M. Randeria, T. Sato, R. Gatt,
T. Takahashi, and K. Kadowaki, Phys. Rev. Lett. {\bf 84}, 4449 (2000).
\bibitem{M.Vishik}
M. Vishik, Rep. Prog. Phys. {\bf 81} 062501 (2018).
\bibitem{shortrange}
B. Keimer, S. A. Kivelson, M. R. Norman, S. Uchida,
and J. Zaanen, Nature {\bf 518}, 179-186 (2015).

\bibitem{site-dependence}
S. Kawasaki, Z. Li, M. Kitahashi
, C.T. Lin, P.L. Kuhns, A.P. Reyes, and  Guo-qing Zheng, Nature {\bf 8}, 1267 (2017).

\bibitem{S-Moriya} T. Moriya and K. Ueda,  Adv. Phys. {\bf 49}, 555 (2000).
\bibitem{Shiba}
H. Shiba, PTP. {\bf 54}, 967 (1975).
\bibitem{K.Yamada}
K. Yamada, Electron Correlation in Metals (Cambridge University
Press, Cambridge, U.K., 2004).



\bibitem{D.J.Scalapino}
D. J. Scalapino, Phys. Rep. {\bf 250}, 329 (1995).
\bibitem{D.J.Scalapino1986}
D. J. Scalapino, E. Loh, and J. E. Hirsch, Phys. Rev. B {\bf 34}, 8190 (1986).

\end{thebibliography}

\begin{thebibliography}{99}
\bibitem{FLEX}
N. E. Bickers and S. R. White, Phys. Rev. B {\bf 43}, 8044 (1991).
\bibitem{A.J.Achkar}
A. J. Achkar, F. He, R. Sutarto, Christopher McMahon, M. Zwiebler, M. Hücker, G. D. Gu, Ruixing Liang, D. A. Bonn, W. N. Hardy, J. Geck and D. G. Hawthorn,  Nat. Mater. {\bf 15}, 616-620 (2016).
\end{thebibliography}
\end{document}